\documentclass[final,3p,times]{elsarticle}
\usepackage{graphicx}  
\usepackage{bm}        
\usepackage{amsmath}
\usepackage{amssymb}
\usepackage[colorlinks,linkcolor=blue,citecolor=blue,urlcolor=blue]{hyperref}
\usepackage[all]{hypcap}
\usepackage{bbold}
\usepackage{subfigure}
\usepackage[title]{appendix}

\biboptions{numbers,sort&compress}
\begin{document}
\begin{frontmatter}
\title{Effect of symmetry defect on the edge resonance of semi-infinite FGM plates}


\author[adress_1,adress_2]{Shengyu Tang}
\author[adress_2]{Wenping Bi\corref{mycorrespondingauthor}}
\ead{wenping.bi@univ-lemans.fr}
\author[adress_1,adress_3,adress_4]{Jingwei Yin}
\author[adress_2]{Vincent Pagneux}

\cortext[mycorrespondingauthor]{Corresponding author. Laboratoire d'Acoustique de l'Universit du Mans (LAUM), UMR 6613, France}

\address[adress_1]{College of Underwater Acoustic Engineering, Harbin Engineering University, Harbin 150001,China}
\address[adress_2]{Laboratoire d'Acoustique de l'Universit du Mans (LAUM), UMR 6613, France}
\address[adress_3]{Acoustic Science and Technology Laboratory, Harbin Engineering University, Harbin 150001,China}
\address[adress_4]{Key Laboratory of Marine Information Acquisition and Security (Harbin Engineering University), Ministry of Industry and Information Technology; Harbin 150001,China}

\begin{abstract}
The effect of asymmetric functionally graded material on the edge resonance and the Fano resonance in semi-infinite FGM plates are reported in this work. The edge resonance is weakened by the material perturbation and the complete mode conversion is illustrated with incident $S_0$ mode. The Fano resonance occurs on the reflected $A_0$ mode as a result of interference between the resonance and direct scattering with incident $A_0$ mode. A hybrid analytical model based on the collocation discretization and the modal decomposition of the elastic field is developed to analyze the scattering properties of the semi-infinite plates. The Fano line-shape is discussed in detail. The results show that the Fano line shape is sensitive to the volume fraction, which is beneficial for the quantitative guided wave application.  
\end{abstract}

\begin{keyword}
Lamb wave \sep Functionally Graded Material \sep Edge resonance \sep Fano resonance
\end{keyword}

\end{frontmatter}


\section{Introduction}

Functionally graded materials (FGM) are characterized by the gradual variation in composition and structure over volume, resulting in corresponding changes in the properties of the material, which are widely used in human-made structures. Specific mixing material and volume fraction are designed to gain more performance \cite{chiuOnedimensionalWavePropagation1999a,hanQuadraticLayerElement2000,liuDispersionWavesCharacteristic2003,kiebackProcessingTechniquesFunctionally2003,vlasieGuidedModesPlane2004,binWavePropagationNonhomogeneous2008,udupaFunctionallyGradedComposite2014,liuFunctionalGradientsHeterogeneities2017} in such as aerospace, medical treatment, and energy equipment, etc. Nowadays, FGM is still an active research area \cite{zhaoFreeVibrationAnalysis2009,baronPropagationElasticWaves2010,berezovskiNumericalSimulationTwodimensional2003,caoCalculationPropagationProperties2011,gravenkampNumericalApproachComputation2012,hebazSemianalyticalDiscontinuousGalerkin2018,yangNewBoundaryElement2020,hedayatrasaNumericalModelingWave2014,tofeldtZerogroupVelocityModes2017} and the difficulty is to improve the forming processes so that the target gradient is achieved with precision \cite{kiebackProcessingTechniquesFunctionally2003,udupaFunctionallyGradedComposite2014,zhangAdditiveManufacturingFunctionally2019}, as well as to uncover the complex nature of fracture mechanics due to material inhomogeneity.

The edge resonance phenomenon has been studied for decades \cite{torvikReflectionWaveTrains1967,wilkie-chancellierNumericalDescriptionEdge2005,pagneuxRevisitingEdgeResonance2006,zernovv.EigenvalueSemiinfiniteElastic2006,ratasseppEdgeResonanceSemiinfinite2008,maurelMultimodalMethodScattering2019} due to the prospective application of Lamb wave in Non-Destructive Testing (NDT). The relative topics about edge resonance on elastic structures could be found in the review article \cite{lawrieEdgeWavesResonance2012} and references therein. The related phenomenon of edge waves is also of intrinsic interest along with other guided waves, which has potential applications in the measurement of material properties. At homogeneous semi-infinite elastic, the influence of symmetry defect has been studied in plates with the beveled edge \cite{wilkie-chancellierNumericalDescriptionEdge2005} or thin pipe with variable curvature \cite{ratasseppEdgeResonanceSemiinfinite2008}, which suggests the edge resonance to persist with the lowest symmetric mode incidence. Interestingly, at a specific parameter, the incident symmetric mode is completely converted into antisymmetric ones. In FGM plates, several works are reported to study the Lamb waves for the character analysis of waves propagating \cite{berezovskiNumericalSimulationTwodimensional2003,caoCalculationPropagationProperties2011,gravenkampNumericalApproachComputation2012,hebazSemianalyticalDiscontinuousGalerkin2018,yangNewBoundaryElement2020}, while the influence of FGM on the edge resonance is still a significant topic to investigate. In the FGM with asymmetrical through-thickness variation, the conventional symmetric and antisymmetric Lamb families are coupled to each other due to symmetry broken, resulting in the complex conversion and interference between modes when a guided Lamb wave reflected at a discontinuity (in this paper a vertical edge). 

Fano resonance \cite{fanoEffectsConfigurationInteraction1961} with asymmetric line shape is a generic phenomenon in scattering problems with multiple resonant pathways and widely exist in harmonic oscillator systems acoustical waveguides, plasmonic nanostructures, electron waveguides photonic or phononic crystals, and metamaterials \cite{nockelResonanceLineShapes1994,joeClassicalAnalogyFano2006,heinFanoResonancesAcoustics2010,lukyanchukFanoResonancePlasmonic2010,xiongFanoResonanceScatterings2016}. It can be theoretically interpreted as the interference between trapped or quasi-trapped modes with the scattering background. The generated asymmetric line shape can be applied in a variety of fields such as beam filters, sensors, and signal communications \cite{papasimakisMetamaterialInducedTransparencySharp2009,lodewijksTuningFanoResonance2013}. The trapped or quasi-trapped modes related to edge resonance are proved existing at homogeneous semi-infinite elastic plates. As the symmetry defect, the weakened edge resonance couple with the scattering path, resulting in the Fano line shape in $A_0$ mode, which allows us to extract rich features for more optimized and quantitative guided wave applications.

In this paper, based on the Chebyshev spectral method, we propose a numerical procedure for studying the Lamb wave propagation in infinite FGM plates. Then the scattering problem is studied based on the modal decomposition. A toy model of FGM is firstly introduced to analyze the effect of material asymmetric on the edge resonance with $S_0$ mode incidence. The Fano lineshape resulting from the interference of resonance and direct scattering is discussed with $A_0$ mode incidence. The results show that we can control the Fano line shape by adjusting the material variation properties. Finally, the Fano resonance is discussed in Cr-Ceramic FGM plates with various volume functions.

\section{Formulation}
\label{sec: formulation}

\begin{figure}[hbt]\centering
  \includegraphics[scale=0.25]{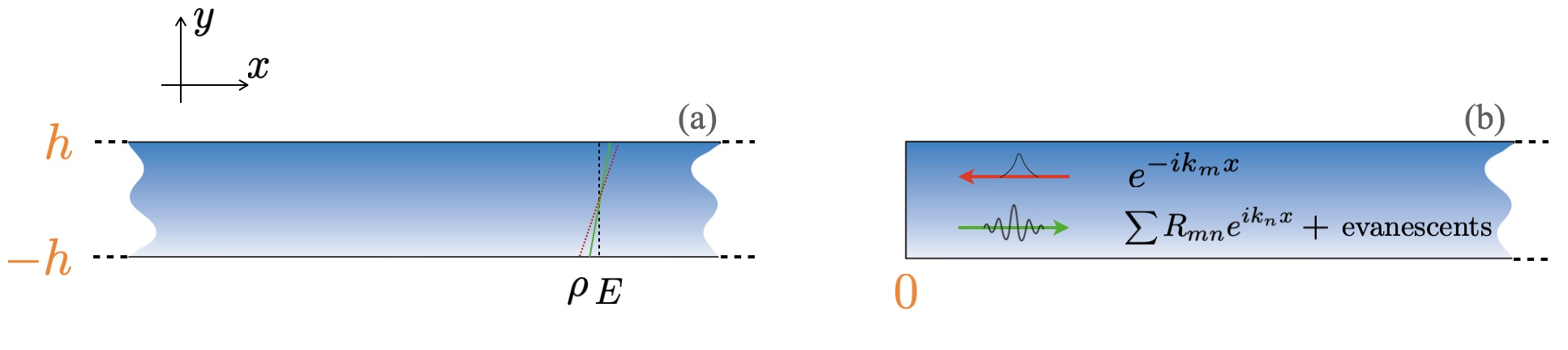}
  \caption{Scheme of (a) infinite FGM plates, and (b) the scattering problem in semi-infinite FGM plates.}
  \label{fig:FGM_configuration}
\end{figure}

In this first section, we introduce the formulation of Lamb wave propagating property and the scattering problem in semi-infinite FGM plates with thickness $2h$ and omit the harmonic time dependence $e^{-i\omega t}$. The frequencies under study are where the lowest three modes propagating.

\subsection{Lamb wave in FGM plates}
There are serval mechanical models for the FGM according to the production process. One example is the interpenetration of two polymer plates, of which the longitudinal and transversal velocities are gradually and symmetrically variable with the thickness while assuming a constant mass density \cite{vlasieGuidedModesPlane2004}. Another example is a plate consists of two components where the volume fractions of the materials vary continuously from one surface to the other \cite{zhaoFreeVibrationAnalysis2009, hebazSemianalyticalDiscontinuousGalerkin2018,caoCalculationPropagationProperties2011}, of which the mass density, Young's modulus, and Poisson's ratio are assumed to be the effective material properties. Furthermore, variable stiffness tensor with constant Poisson's ratio \cite{baronPropagationElasticWaves2010,yangNewBoundaryElement2020} and specified  velocity function with constant density \cite{tofeldtZerogroupVelocityModes2017} are also studied.

Since we are considering the edge scattering of inhomogeneous and non-symmetric continuous variation plates. It is reasonable to firstly study a toy model with the constant Poisson's ratio $\nu$ and linearly varying mass density as well as Young's modulus (see Fig. \ref{fig:FGM_configuration}(a)). The functions of volume fractions are assumed as
\begin{equation}
	\begin{aligned}
         	\rho(y)&=\rho_0 (1+py/h),\\
           E(y)&=E_0 (1+py/h),
         \end{aligned}
\end{equation}
where $p\in[0,1)$ is the material gradient, $\rho_0=2700 \ kg/m^3$, $E_0=69\ \text{GPa}$ are respectively the reference material properties of the homogenous plates used in the following study. The equations of motion in FGM plates \cite{vlasieGuidedModesPlane2004} are

\begin{equation}
	\begin{aligned}
		-\rho \omega^2 \mathbf{w}= (\lambda+\mu)&\nabla(\nabla \cdot \mathbf{w})+\mu \nabla^2 \mathbf{w}\\
   &+\partial_y \lambda(\nabla \cdot \mathbf{w})\textbf{n}_y + \partial_y \mu (\nabla \textbf{w}+{^{t}\nabla} \textbf{w})\textbf{n}_y,
	\end{aligned} 	
\end{equation}
and the stress-free boundary conditions are
\begin{equation}
	\sigma_{xy}=\sigma_{yy}=0, \quad \text{on}\ y=\pm h,	
\end{equation}
where $\mathbf{w}=(u_x,u_y)^{T}$ is the displacement and $\mathbf{\sigma}$ is the stress tensor with
\begin{equation}
	\begin{aligned}
		\sigma_{xx} &= \lambda\partial_y u_y+(\lambda+2\mu)\partial_x u_x,\\
		\sigma_{xy} &= \mu(\partial_y u_x+\partial_x u_y),\ \ \sigma_{yy} = (\lambda+2\mu)\partial_y u_y+\lambda\partial_x u_x.
	\end{aligned}
\end{equation}
The boundary conditions on the free edge of a semi-infinite plate read as
\begin{equation}
	\sigma_{xx}=\sigma_{xy}=0,\quad \text{on} \ x=0.
	\label{eq:edge bc}
\end{equation}
The dimensionless frequency is defined as $\Omega = \omega h/c_0$, with the reference velocity $c_0=1000\pi \ \text{m/s}$. 

\begin{figure}[hbt]\centering
  \includegraphics[scale=0.8]{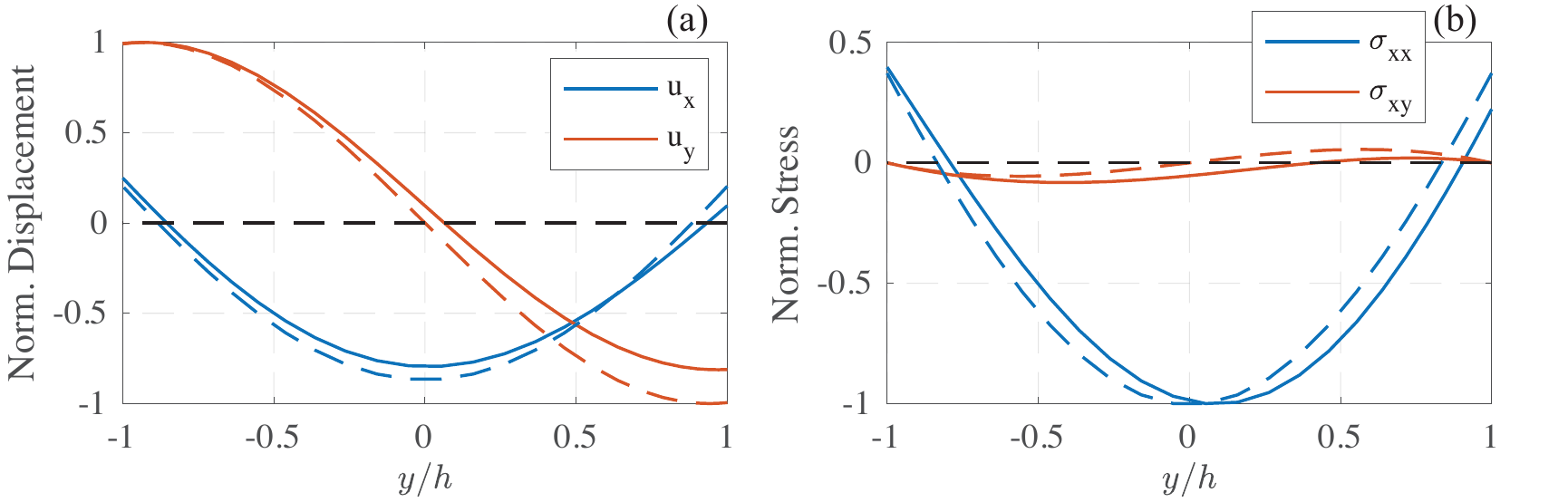}
  \caption{The effect of FGM on the $S_0$ mode profiles at $\nu=0.33$, $p=0.185$, $\Omega=2.335$. The displacement (a) and stress (b) profiles are normalized to the largest amplitude. Dashed lines are the homogeneous plates.}
  \label{fig: mode profile}
\end{figure}

We develop the configuration reported in reference \cite{pagneuxRevisitingEdgeResonance2006}, proposing a procedure based on the Chebyshev spectral method for computing the Lamb models in FGM plates (detail in \ref{app: Numerical determination of Lamb mode in FGM plates}). The Lamb mode can be found by separating the variables in the form
\begin{equation}
	\begin{pmatrix}
		\mathbf{\hat{X}}(x,y)\\\mathbf{\hat{Y}}(x,y)
	\end{pmatrix}=e^{ikx}\begin{pmatrix}
		\mathbf{X}(y)\\\mathbf{Y}(y)
	\end{pmatrix},
\end{equation}
where $\mathbf{X}=(u_x,\sigma_{xy})^{T}$, $\mathbf{Y}=(-\sigma_{xx},u_y)$, $k$ corresponds to the wavenumber. The spectrum and profile of Lamb modes then can be obtained by solving the eigenvalue problem. The obtained results are verified by the power series technique \cite{caoCalculationPropagationProperties2011} and the Scaled Boundary Finite Element Method (SBFEM) \cite{gravenkampNumericalApproachComputation2012}. The corresponding mode direction is determined by the imaginary part of evanescent modes (complex or imaginary wavenumber) or the group velocity of propagating modes (real wavenumber) \cite{pagneuxLambWavePropagation2002}. The influence of FGM on the mode profile is plotted in Fig. \ref{fig: mode profile}. Obviously, the displacements and stresses are no more symmetry with respect to the mid-plane, but introduced a perturbation. Based on this fact, they are quasi-symmetric or quasi-antisymmetric modes compared to the homogeneous case. In the following study, we omit the quasi terms for simplicity.

 \subsection{Reflection coefficient}
 The reflection coefficients on the free end of a semi-infinite FGM plate are defined as
  \begin{equation}
 	R_{mn},  \quad m,n=A_0,\ S_0,\ A_1\ \cdots.
 \end{equation}
 where the subscript $m$ and $n$ represent the incident and the reflected modes. In order to describe the reflection process, we introduce the following scattering solutions 
\begin{equation}
	\mathbf{\phi}(x,y) = e^{ik_{m}^{-}x}\mathbf{\Phi}_{m}^{-}(y)+\sum_{n=1}^{N}R_{m n} e^{ik_{n}^{+}x}\mathbf{\Phi}^{+}_n(y),
	\label{eq:scattering solutions}
\end{equation}
where $\phi=(u_x,u_y,\sigma_{xx}, \sigma_{xy})^T$, $k_{m}^{-}$ corresponds to the wavenumber of incident left propagating mode, $k_{n}^{+}$ corresponds to the wavenumber of reflected right going modes (propagating and evanescent), $R_{m n}$ is the reflection coefficient of modes due to the incident on the free edge, $\mathbf{\Phi}$ is the corresponding mode profile and $N$ is the truncated number which should be sufficient enough to meet the boundary conditions on the edge in numerical calculating \cite{torvikReflectionWaveTrains1967}. Submitting the boundary conditions Eq.(\ref{eq:edge bc}) into Eq.(\ref{eq:scattering solutions}), then we obtain the reflection coefficient of each mode. 
 
 At homogeneous semi-infinite plates, the symmetry indicates that the coupling between symmetric and antisymmetric cases is not possible. However, it appears due to the symmetry defect in FGM plates. Consider the propagating modes, the reflection matrix is in the form
 \begin{equation}
 	\mathbf{R}=\begin{pmatrix}
 		R_{A_0 A_0} & R_{A_0 S_0} & R_{A_0 A_1}\\
 		R_{S_0 A_0} & R_{S_0 S_0} & R_{S_0 A_1}\\
 		R_{A_1 A_0} & R_{A_1 S_0} & R_{A_1 A_1}
 	\end{pmatrix},
 \end{equation}
 where the coupling elements $R_{S A}$ and $R_{A S}$ are zero at homogeneous plates.

\subsection{Quality factor}
The edge resonance is studied from the point of view of complex resonance in reference \cite{pagneuxRevisitingEdgeResonance2006} and it is reported that only one complex resonance frequency exists for each Poisson's ratio $\nu \in [0,0.5)$. We define the quality factor of the edge resonance in semi-infinite plates as
\begin{equation}
	Q = \frac{\text{Re}(\Omega_R)}{2|\text{Im}(\Omega_R)|}.
	\label{eq: quality factor}
\end{equation}
This is proportional to the ratio of stored energy over energy loss per cycle. The temporal dependence 
\begin{equation}
	e^{-i\Omega_R t} = e^{-i\text{Re}(\Omega_R)t}e^{\text{Im}(\Omega_R)t},
\end{equation}
leads to $\text{Im}(\Omega_R)<0$ which corresponds to the time decay (ring time) of the complex resonance. The complex number with $\text{Im}(\Omega_R)<0$ corresponds to the quasi-trapped mode behaving in a quasi-resonance (finite quality factor) near the resonance frequency, and $\text{Im}(\Omega_R)=0$ corresponds to the trapped mode (infinite quality factor). Two real resonances have been found for the real frequency at $\nu_1 = 0$ and $\nu_2 \approx 0.2248$ and, are inferred owing to the hidden symmetry of the scattering problem which results in the decoupling between propagating and evanescent waves \cite{pagneuxRevisitingEdgeResonance2006,zernovv.EigenvalueSemiinfiniteElastic2006}. 
   
 \subsection{Energy conservation}
 The bi-orthogonality relation of layered elastic plates has been proved by Murphy et al \cite{murphyOrthogonalityRelationRayleigh1994}, which is the same form as the one of homogeneous plates. Thus, we could calculate the time-averaged energy flux per unit width passing through any arbitrary position $x$ with
\begin{equation}
	\Pi = \frac{-i\omega}{4} \int_{-h}^{h}(\sigma_{xx} \bar{u}_x+\sigma_{xy} \bar{u}_y-\bar{\sigma}_{xx} u_x-\bar{\sigma}_{xy} u_y) \text{dy},
\end{equation}
where the over-line represents conjugate. In the harmonic regime, the conservation of the complete energy reduces to the conservation of the time-averaged energy flux \cite{maurelMultimodalMethodScattering2019}. The energy flux of plane waves conserved across the section reads as the equality between the incident energy flux and the reflected energy flux.
\begin{equation}
	\sum \Pi_m = \sum \Pi_n
\end{equation}
The following numerical results are verified by this energy balance.

\section{Edge reflection of FGM plates}
\label{sec: edge reflection of FGM plates}

In this section, we explain how the material symmetry defect influencing the edge resonance, and present the Fano resonance in semi-infinite FGM plates for a given semi-infinite plate with FGM. Two reflection problems are presented (see Fig. \ref{fig:FGM_configuration}(b)). The first one is the reflection problem with the incident $S_0$ mode, where the influence of FGM on the edge resonance is studied and the complete mode conversion occurs. The second one is the reflection problem with the incident $A_0$ mode, where the Fano resonance phenomenon is observed and discussed.  

\subsection{Effect on the edge resonance ($S_0$ incident)}
\label{subsec: Effect on the edge resonance}

\begin{figure}[hbt]\centering
  \includegraphics[scale=0.8]{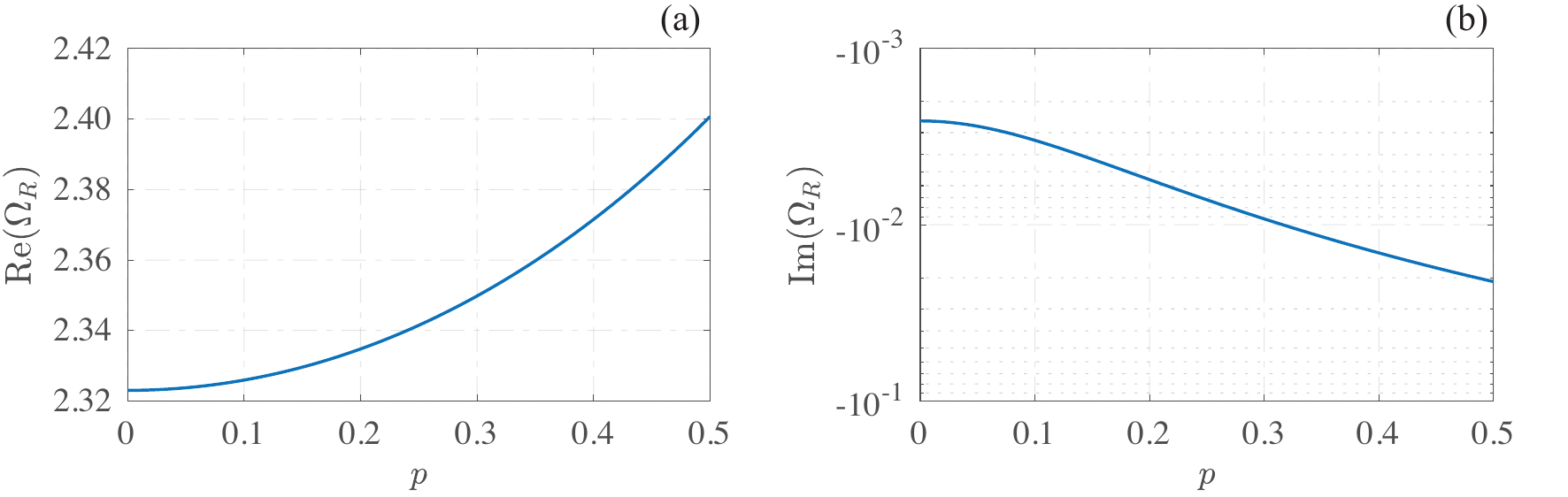}
  \caption{The effect of FGM on the complex resonance frequency. (a) real part and (b) imaginary part. The Poisson's ratio is $\nu=0.33$, and the reference material properties are $\rho_0=2700 \ kg/m^3$, $E_0=69\ \text{GPa}$. }
  \label{fig: effect on complex resonance freq.}
\end{figure}

\begin{figure}[hbt]\centering
  \includegraphics[scale=0.8]{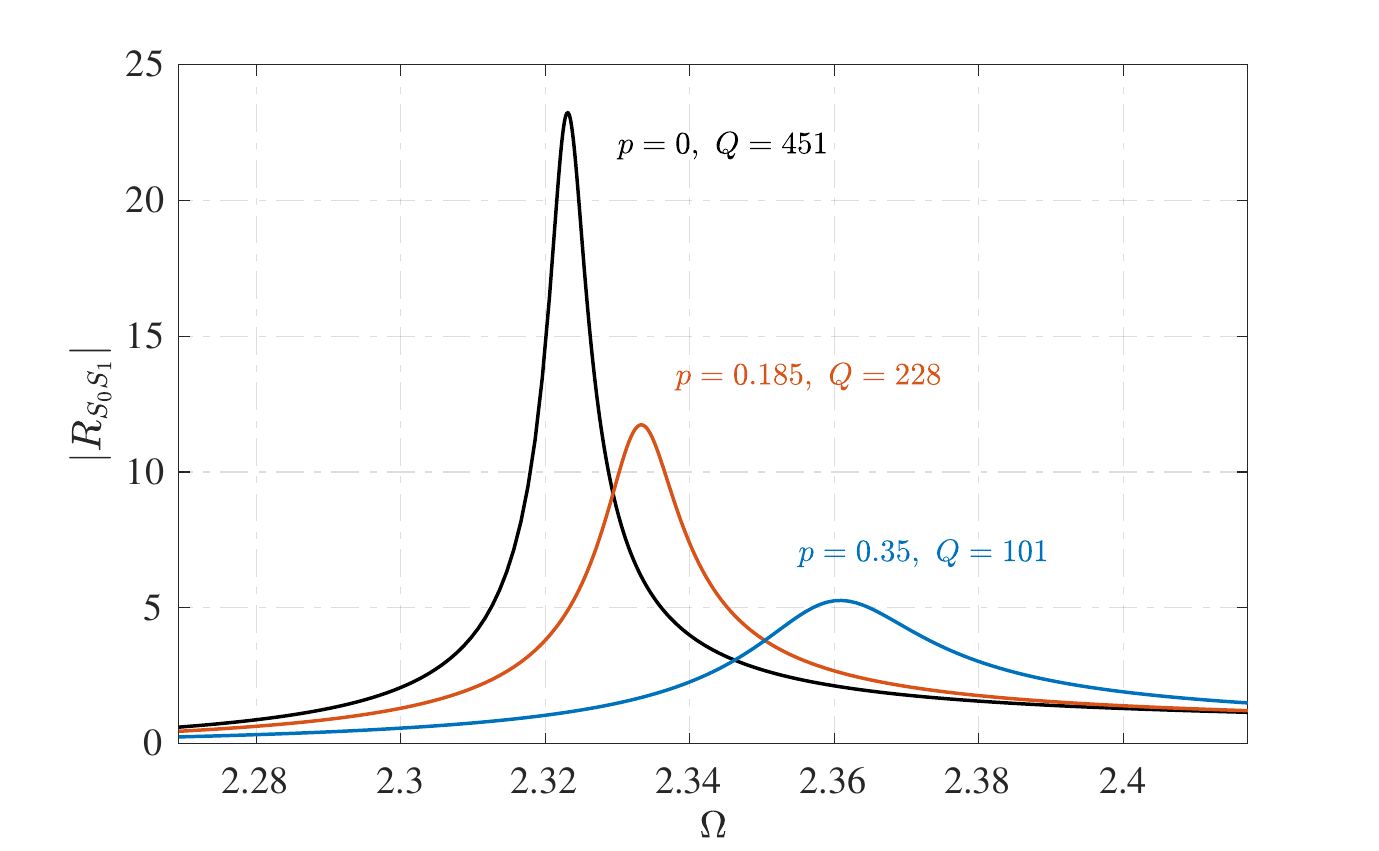}
  \caption{The reflection coefficient spectrum of evanescent mode $S_1$ near the resonance frequency. The incident mode is $S_0$ at $\nu=0.33$. The resonance quality factor $Q$ is calculated with Eq.(\ref{eq: quality factor}).}
  \label{fig: S1 spectrum}
\end{figure}

The effect of material asymmetry on the edge resonance is studied with the incident $S_0$ mode. In geometry asymmetric system, such as the semi-infinite plate with the beveled free edge \cite{wilkie-chancellierNumericalDescriptionEdge2005}, as well as the semi-infinite pipe with variable curvature \cite{ratasseppEdgeResonanceSemiinfinite2008}, the edge resonance is suggested to persist even there is a coupling between the symmetric and antisymmetric modes. In this paper, the asymmetric is produced by the FGM. 

The edge resonance frequency of the semi-infinite FGM plate could be found by two methods. The one is searching the pole of the reflection coefficient $R_{S_0 S_0}$ in the complex frequency plane from the point of view of complex resonance theory, and the real part of this pole is the edge resonance frequency. Another is finding the maximum of fist complex evanescent $S_1$ mode in the spectrum. Both of them could be carried out following the process mentioned in Sec. \ref{sec: formulation}, concerning the reflection coefficient. In Fig. \ref{fig: effect on complex resonance freq.}, the real and imaginary parts of the complex resonance frequency are studied as a function of material gradient $p$. A rapid increase in the real part of the complex resonance frequency while a decrease for the imaginary part is observed as rising the gradient, which means the edge resonance shifts toward high frequencies, accompanied by a decrease in resonance intensity. This phenomenon could be clearly seen in the spectrum of the evanescent $S_1$ mode (see Fig. \ref{fig: S1 spectrum}).

\begin{figure}[hbt]\centering
\subfigure{
  \includegraphics[scale=0.8]{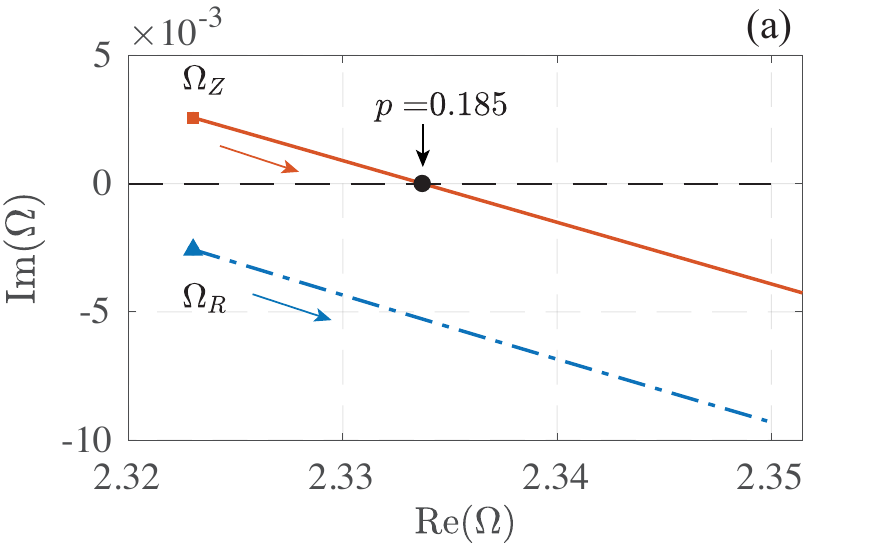}}
  \subfigure{
  \includegraphics[scale=0.8]{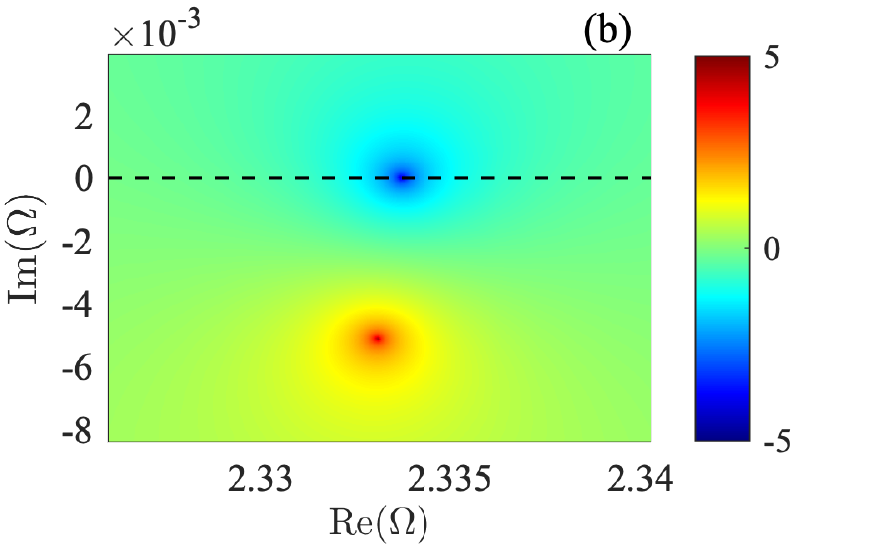}}
  \caption{(a) The pole (dot-dashed line) and zero (solid line) of the $S_0$ reflection coefficient as material gradient increases. The incident is $S_0$ mode at $\nu=0.33$. The square and triangle correspond to the case of homogeneous plates ($p=0$) respectively. (b) The reflection coefficient $\text{log}_{10}(|R_{A_0 A_0}|)$ in the complex frequency plane at $p=0.185$.}
  \label{fig: s0 shift}
\end{figure}

\begin{figure}[hbt]\centering
  \includegraphics[scale=0.8]{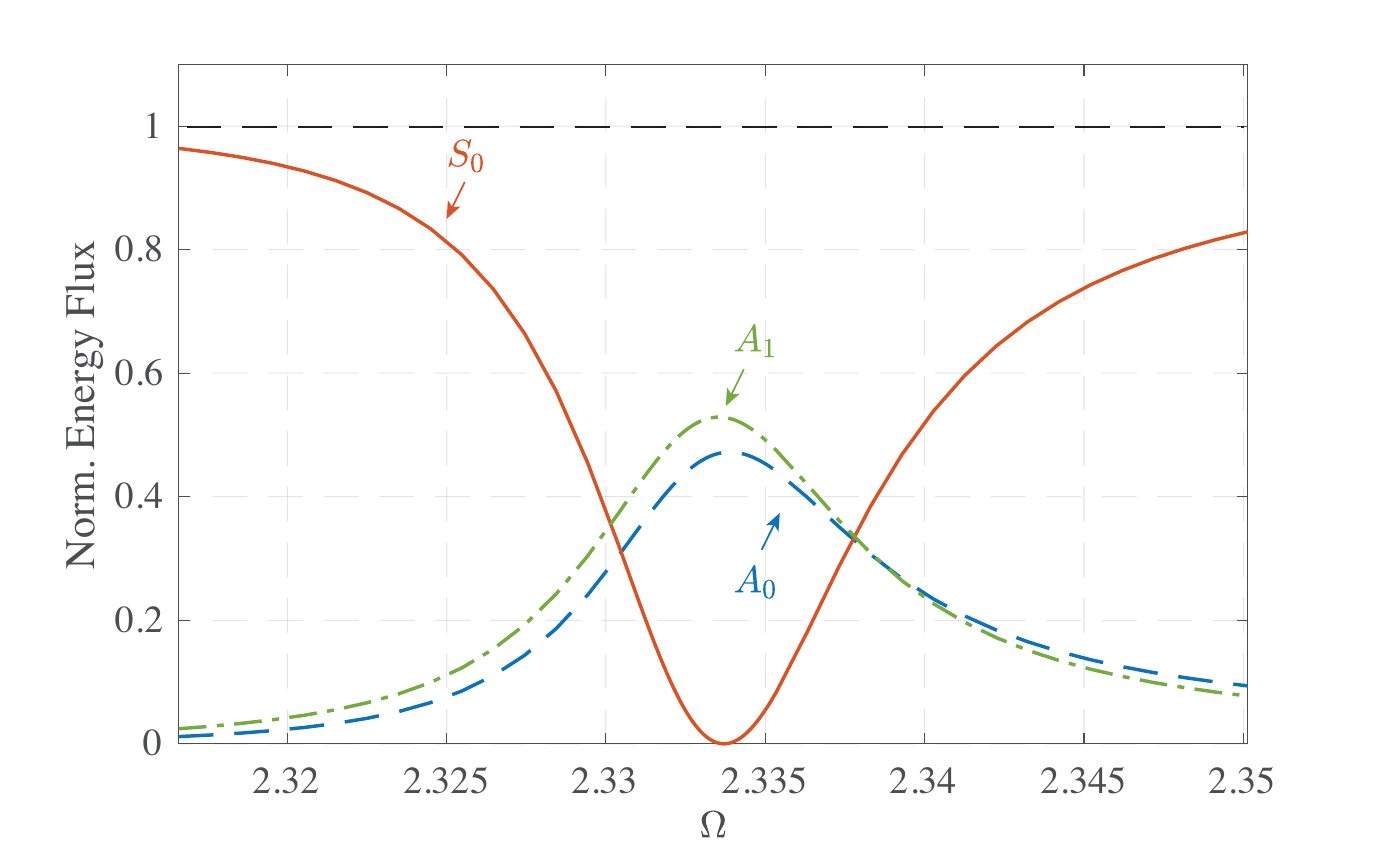}
  \caption{Complete mode conversion at $\nu=0.33$, $p=0.185$ with incident the $S_0$ mode. The black dashed line corresponds to the reflected $S_0$ mode of homogeneous plates.}
  \label{fig: completely conversion}
\end{figure}

Besides the pole of reflection coefficient $R_{S_0 S_0}$, it is interesting to report the behavior of the zero $\Omega_Z$ in the complex plane. In a symmetrical system, the poles and zeros are conjugated with respect to the real frequency axis. By varying the material gradient, the symmetry is broken and mode coupling between the quasi-symmetric and quasi-antisymmetric modes becomes possible. When increasing the material gradient, the zeros shift at the same laws as the poles (as depicted in Fig. \ref{fig: s0 shift} (a)). It is remarkable that not only are they not conjugate, but the real parts of them also have a deviation. In particular, the trajectory of the zero intersects with the real frequency axis, which means the energy of the incident $S_0$ mode is complete converted to the quasi-antisymmetric ones. We show the spectrum of $R_{S_0 S_0}$ in Fig. \ref{fig: s0 shift} (b), and report the complete mode conversion by investigating the normalized reflection energy flux of each mode in Fig. \ref{fig: completely conversion}, where the energy ratio of the $S_0$ mode is reduced to zero at $p=0.185$.

\subsection{Fano resonance ($A_0$ incident)}
\label{subsec: Fano resonance}

\begin{figure}[hbt]\centering
  \includegraphics[scale=0.8]{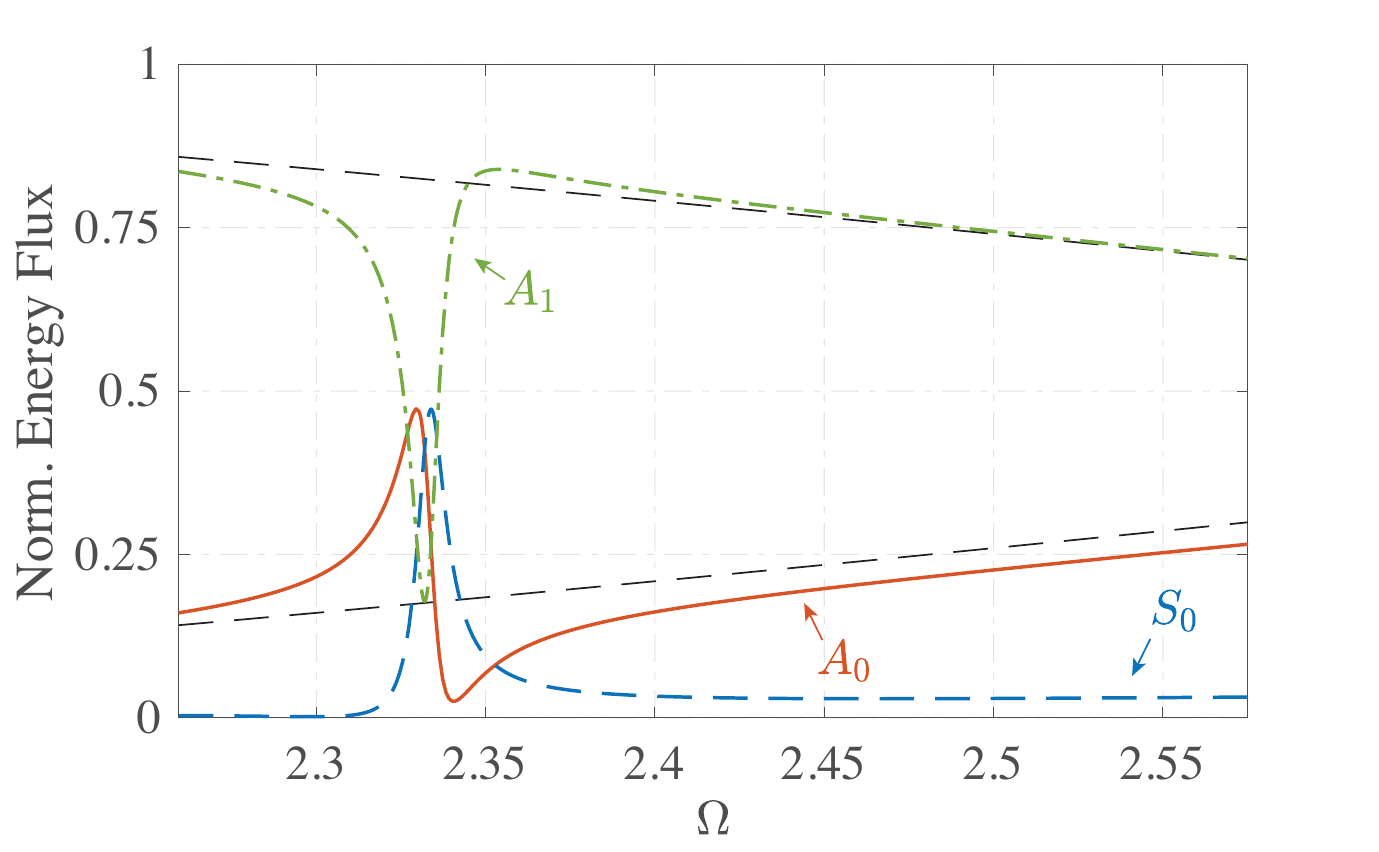}
  \caption{The Fano resonance in semi-infinite FGM plates at $\nu=0.33$, $p=0.185$. The incident is quasi-antisymmetric $A_0$ mode. The black dashed line corresponds to the homogeneous plates ($p=0$); The red solid line corresponds to $A_0$, the blue dashed line corresponds to $S_0$ and the green dash-dot line corresponds to $A_1$.}
  \label{fig: FanoExample}
\end{figure}

We know that without the symmetry defect, $A_0$ and $A_1$ will convert energy to each other while the symmetric modes will not participate. This conversion is very gentle on the spectrum (as illusulated in Fig. \ref{fig: FanoExample}). In the case of antisymmetric mode incidence, the edge resonance is not capable to be excited. However, as the symmetry broken (due to the material variation in the following), the situation changes because of the mode coupling on the free edge. It is then interesting to study the impact of both the material gradient and Poisson's ratio on reflection coefficients.
    
\begin{figure}[hbt]\centering
\subfigure{
  \includegraphics[scale=0.8]{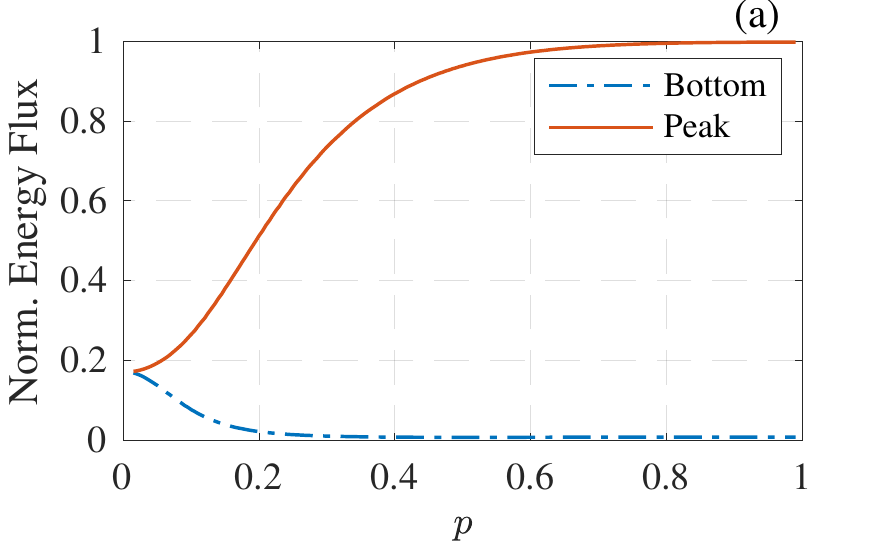}}
  \subfigure{
  \includegraphics[scale=0.8]{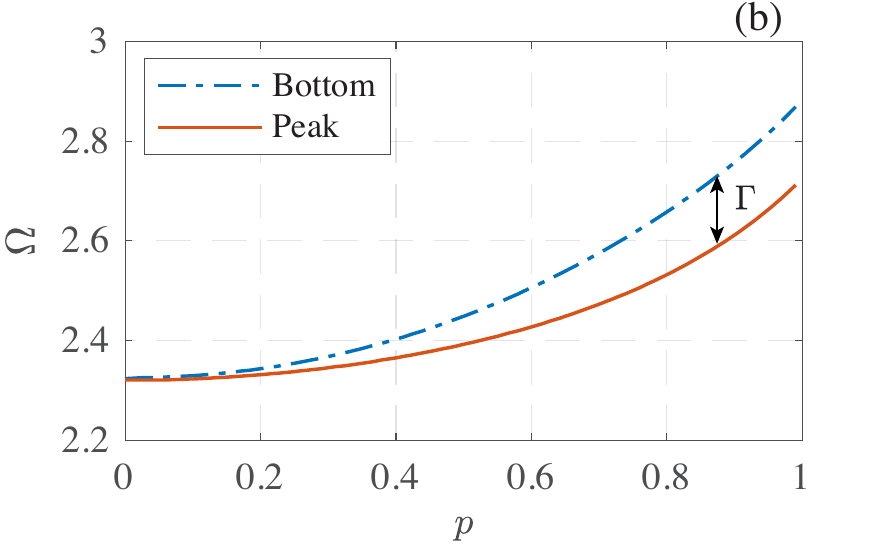}}
  \caption{The Fano resonance as a function of material gradient $p$ at $\nu=0.33$. (a) The bottom's and peak's amplitude of the Fano line, and (b) their corresponding frequencies. $\Gamma$ is the bandwidth of the resonance.}
  \label{fig: Fano vs p}
\end{figure}

We firstly take Poisson's ratio $\nu=0.33$ as an example. When incident the quasi-antisymmetric $A_0$ mode, the reflection coefficient $R_{A_0 A_0}$ behaves a resonance peak near the edge resonance frequency and a reflection zero close to the peak at $p=0.185$ (as shown in Fig. \ref{fig: FanoExample}). This phenomenon is called Fano resonance which is inferred by the coupling between the resonance of quasi-symmetric modes and the direct scattering of quasi-antisymmetric modes. 

Results in Fig. \ref{fig: Fano vs p} show the effect of the FGM gradient on the normalized energy flux of quasi-symmetric $A_0$ mode. The Fano line shape becomes significant when the symmetry broken reaches a certain level. The peak amplitude continues to increase and finally stabilizes to one, while the minimum is to zero. In this process, the resonance frequency increases, which is related to the edge resonance frequency varying in Fig. \ref{fig: effect on complex resonance freq.}(a), and the bandwidth of the Fano resonance continues to increase (see Fig. \ref{fig: Fano vs p}(b)). Indeed, when the material gradient is much larger than $\xi/Q_0$, the resonance degenerates to symmetric Breit–Wigner (or Lorentzian) lineshape \cite{joeClassicalAnalogyFano2006}, where $\xi$ is a constant and $Q_0(\nu)$ is the quality factor of the edge resonance of homogeneous plates. 

\begin{figure}[hbt]\centering
\subfigure{
  \includegraphics[scale=0.8]{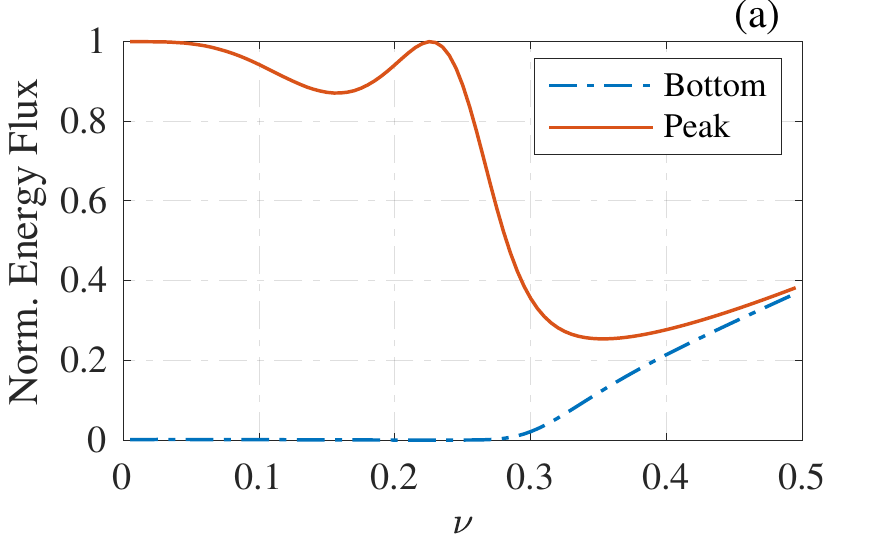}}
  \subfigure{
  \includegraphics[scale=0.8]{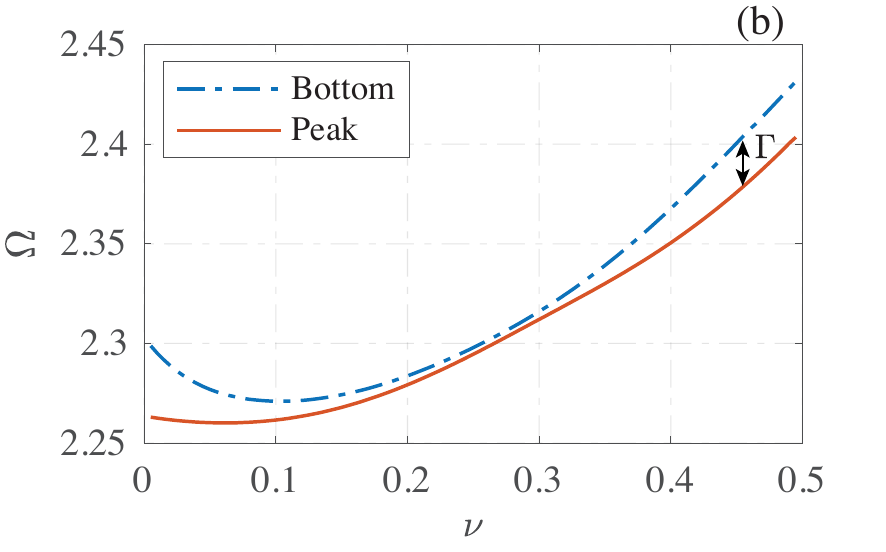}}
  \caption{The Fano resonance as a function of Poisson's ratio $\nu$ at $p=0.1$. (a) The bottom's and peak's amplitude of the Fano line, and (b) their corresponding frequencies. $\Gamma$ is the bandwidth of the resonance. }
  \label{fig: FanoNu}
\end{figure}

The Fano phenomenon is very sensitive to the edge resonance. In the previous section, we illustrated the edge resonance has an infinite quality factor at $\nu=0$ and $\nu \approx 0.2248$, which implies that a tiny material gradient can lead to abrupt influence on the Fano line shape. The relative size of $p$ is a balance to the parameter $\xi/Q_0$. The amplitudes and bandwidth $\Gamma$ of the Fano resonance are presented as a function of Poisson's ratio $\nu$ at $p=0.1$ (see Fig. \ref{fig: FanoNu}). Near the Poisson's ratio where trapped modes exist of homogeneous plates, the peak value of the Fano resonance shows a similar behavior of that $p\approx 1$ at $\nu=0.33$. The bandwidth increases when the Poisson's ratio is farther away from $\nu \approx 0.2248$. We remark that the increase on the left and right of this equilibrium is due to deviations from the parameter $\xi/Q_0$ under the corresponding Poisson’s ratio. The left is far larger while the right is smaller.

\begin{figure}[hbt]\centering
  \includegraphics[scale=0.8]{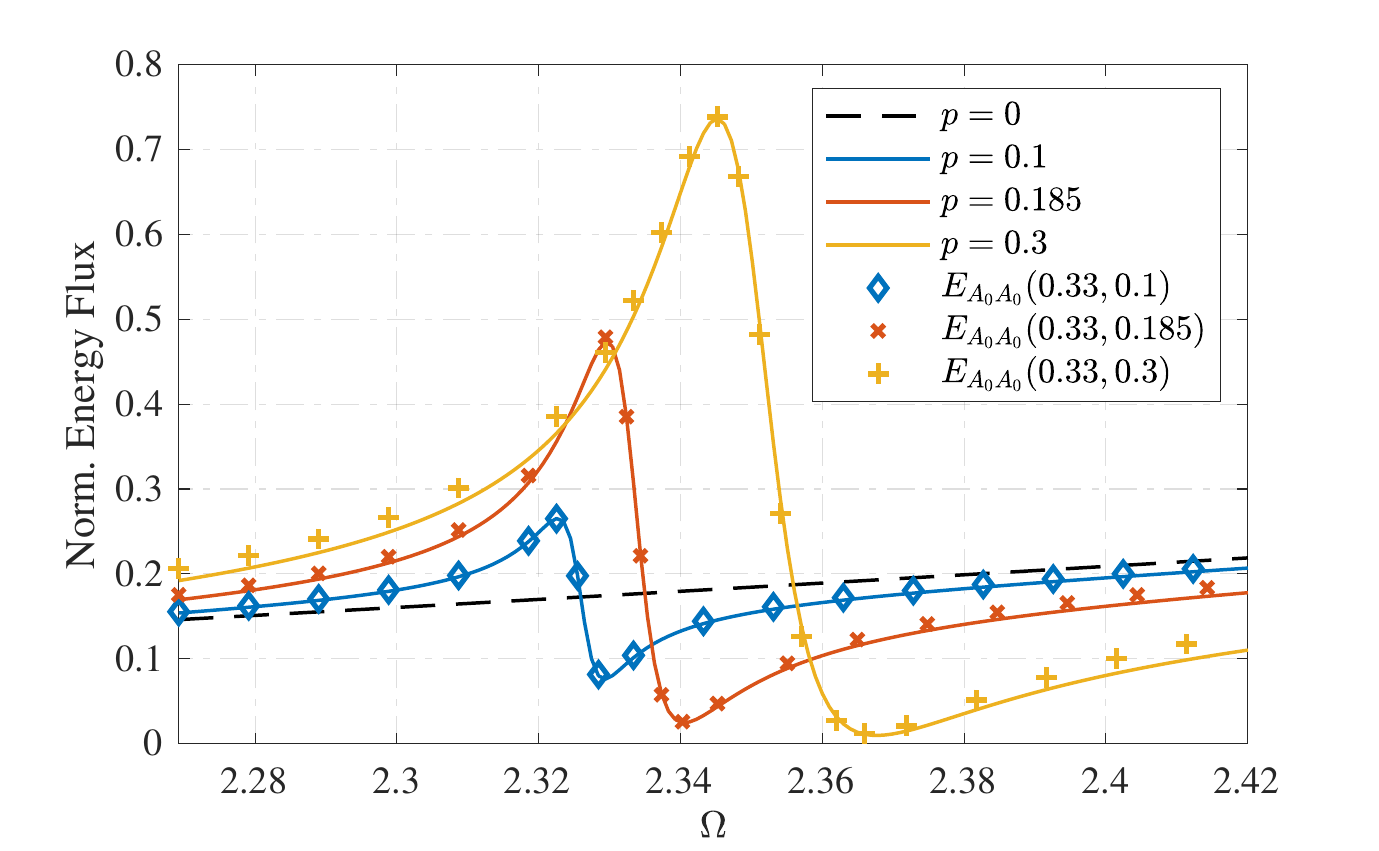} 
  \caption{The empirical formula fitting of normalized reflected $A_0$ energy flux. The Poisson's ratio is $\nu=0.33$. The black dashed line corresponds to the homogeneous plate ($p=0$). The solid lines are the numerical results, and the markers are obtained by Eq.(\ref{eq: Fano equation}) with $\delta=-0.0018 + 0.0028i$, $\delta=-0.0062 + 0.0032i$ and $\delta=-0.0157 + 0.0046i$ for $p=0.1,\ 0.185$ and $0.3$ respectively.}
  \label{fig:}
\end{figure}

The normalized energy flux of reflected $A_0$ can be quantified using \cite{nockelResonanceLineShapes1994}
\begin{equation}
	E_{A_0 A_0}(\nu,p) = |R_{A_0 A_0}|^2\frac{(\Omega-\text{Re}(\Omega_{R})+\delta)^2}{(\Omega-\text{Re}(\Omega_{R}))^2+\Gamma^2},
	\label{eq: Fano equation}
\end{equation}
where $R_{A_0 A_0}$ is the reflection coefficient of homogeneous plates, $\Omega_R(\nu,p)$ is the complex edge resonance frequency of FGM plates, $\delta(\nu, p) = \delta_r +i\delta_i$ determines the asymmetry of the Fano line shape, and $\Gamma=\text{Im}(\Omega_R)$ is the bandwith. It is essential that $\delta_i$ is not zero when $p$ is not much greater than the corresponding parameter $\xi/Q_0$, which determines the bottom amplitude of the Fano resonance. While beyond this region, $\delta_i=0$, corresponding to the behaviour at large material gradient (see example in Fig. \ref{fig: Fano vs p}(a)).

\begin{figure}[hbt]\centering
\subfigure{
  \includegraphics[scale=0.8]{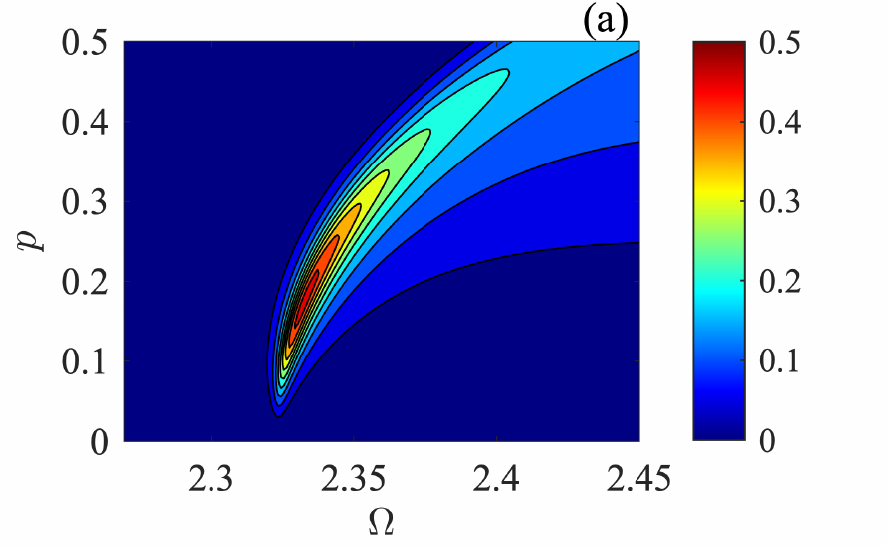}}
\subfigure{
  \includegraphics[scale=0.8]{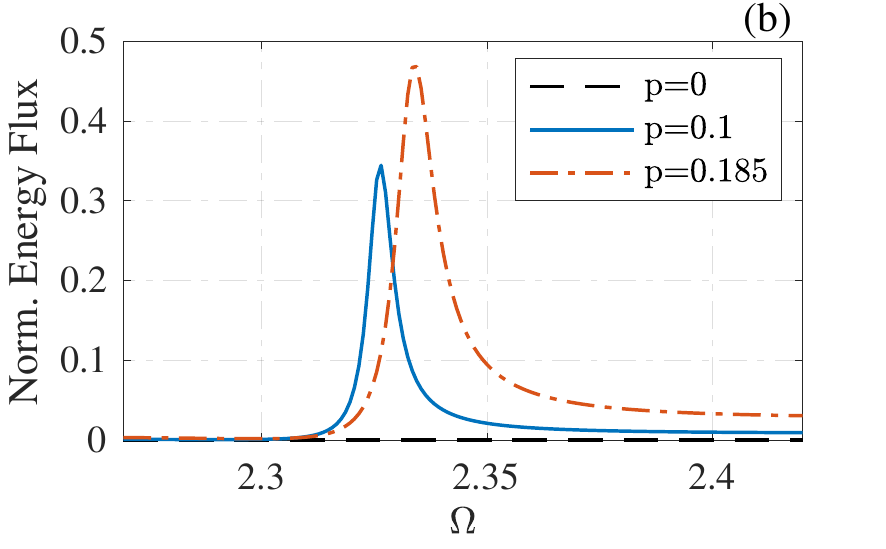}}
  \caption{(a) The normalized energy flux of reflected $S_0$ mode. The incident is $A_0$ mode at $\nu=0.33$; (b) The extract section at $p=0,\ p=0.1$, and $p=0.185$.}
  \label{fig: A0 in S0 energy}
\end{figure}

\subsection{Reflection properties selection}

The reflected symmetric modes are zero as we have mentioned in the previous section of homogeneous plates when antisymmetric mode incidents. The inhomogeneity of material properties leads to the energy conversion from incident $A_0$ mode to the right propagating $S_0$ mode on the free edge of FGM plates. Therefore, we look at the normalized energy flux of $R_{A_0 S_0}$ as a function of material gradient and frequency at $\nu=0.33$ (see Fig. \ref{fig: A0 in S0 energy}(a)). There is a peak of the reflected $S_0$ which approaches $0.5$ of the complete energy. However, we remark that this peak is not able to reach one, which means the incident $A_0$ mode is not able to completely converts to the $S_0$ mode. Since this coupling is caused by symmetry breaking, it is possible to select the reflection properties of the edge by studying the configuration of plates with both material and geometry symmetry defects.

\section{Application example}
\label{sec: application example}

\begin{figure}[hbt]\centering
\subfigure{
  \includegraphics[scale=0.8]{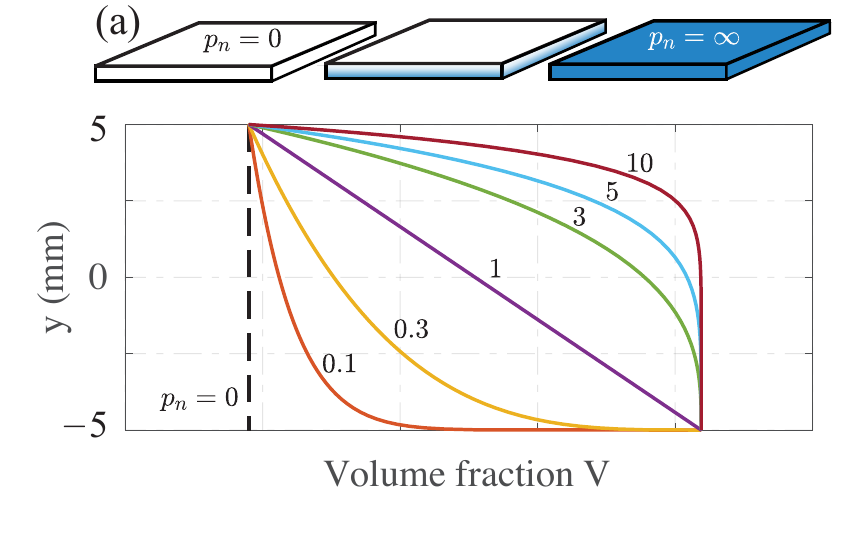}}
\subfigure{
  \includegraphics[scale=0.8]{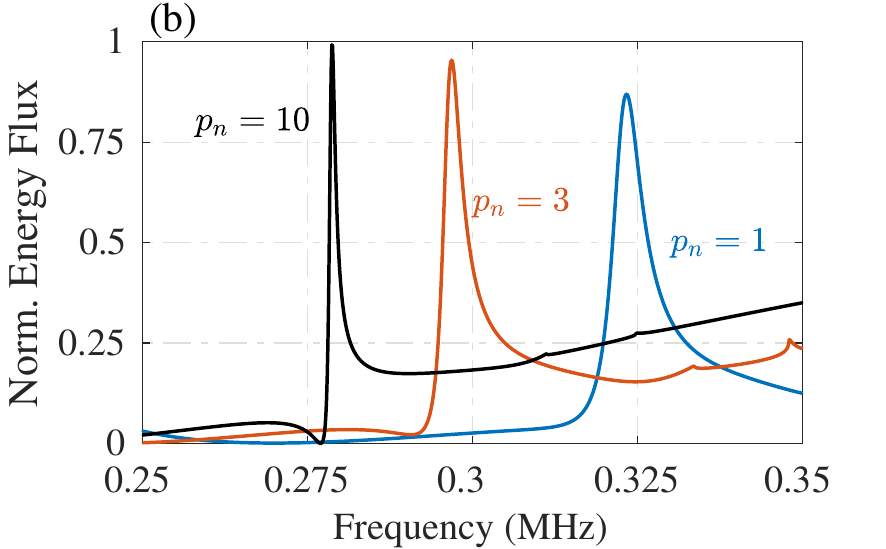}}
  \caption{The Fano resonance in Cr-Ceramic FGM plates. (a) The diagram of volume fraction. $p_n=0$ corresponds to the pure ceramics, $p_n=\infty$ corresponds to the pure metal Cr, and the others are FGM with material properties Continuous varying. (b) The Fano line shapes in Cr-Ceramic FGM plates with different volume fraction.}
  \label{fig: Cr-Ceramic}
\end{figure}

Ceramic-metal-based FGM is usually applied in a high-temperature environment which assumes that one side is pure ceramic, and the other end is pure metal, with a gradient change in the middle. An example is chosen where the FGM plate is consists of metal Cr and ceramics \cite{caoCalculationPropagationProperties2011}. The effective material properties, density as well as two parameters of isotropic solid $\lambda$ and $\mu$, are varying in the power-law through the thickness
\begin{equation}
	V(y) = V_m f_1(y/h)+V_c f_2(y/h)
\end{equation}
where $V_m$ and $V_c$ correspond to the material properties of metal Cr and ceramics respectively
\begin{equation}
\begin{aligned}
	&\rho_m = 7190 \ \text{kg}\ \text{m}^{-3},\quad \lambda_m = 74.2\ \text{GPa},\quad &\mu_m = 102.5\ \text{GPa}\\
	&\rho_c = 3900 \ \text{kg}\ \text{m}^{-3},\quad \lambda_c = 138\ \text{GPa},\quad &\mu_c = 118.11\ \text{GPa}\\
\end{aligned}
\end{equation}
The volume fractions $f_1$ and $f_2$ are
\begin{equation}
\begin{aligned}
	f_1(y/h) &= 1-(\frac{1-y/h}{2})^{p_n}\\
	f_2(y/h) &= (\frac{1-y/h}{2})^{p_n}
\end{aligned}
\end{equation}
with $p_n$ a non-negative from $0$ to infinite. The volume fractions of FGM plates with thickness $2h=1 \text{cm}$ are depicted in Fig. \ref{fig: Cr-Ceramic}(a), where $p_n=0$ corresponds to the pure ceramics, $p_n=\infty$ corresponds to the pure metal Cr, and the others are FGM with material properties Continuous changing.

Considering Lamb $A_0$ mode incidence to the edge of semi-infinite FGM plates and focusing on the reflected $A_0$ mode, we display the variation of Fano line shape concerning the volume fraction in Fig. \ref{fig: Cr-Ceramic}(b). Remarkably, the Fano line shape is sensitive to the volume fraction, which has the potential in material characterization.

\section{Conclusions}
The influence of asymmetric FGM on the edge resonance phenomenon was firstly investigated in semi-infinite plates. The scattering of quasi-symmetric mode $S_0$ at the edge was modeled by applying the modal decomposition technique which allows us to easily estimate the contribution of each mode in the expansion. The accuracy of the computation can be checked by the energy conservation. The study of material inhomogeneity has shown that the edge resonance is weakened and the resonance shift to high frequency. In particularly, at specific material parameters, the incident $S_0$ is completely converted to quasi-antisymmetric $A_0$ and $A_1$.

The Fano resonance was produced resulting in the coupling between the resonance and scattering background when the quasi-antisymmetric $A_0$ incident. This line shape was accurately modeled concerning the complex edge resonance frequency of the FGM plates. The study implied that one can control the Fano resonance of $A_0$ mode by adjusting the material volume fraction of FGM, which allow us to extract rich feature for more quantitative guided wave application, especially in material characterization.

\section*{Acknowledgement} 
This work was supported by the China Scholarship Council [No.201806680019].

\begin{appendix}

\section{Numerical determination of Lamb mode in FGM plates}
\label{app: Numerical determination of Lamb mode in FGM plates}
The numerical procedure follows the one introduced in \cite{pagneuxRevisitingEdgeResonance2006}, where the Chebyshev discretion points and differential matrix are used. We could solve the eigenvalue problem of the dispersion relation of FGM plates 
\begin{equation}
  \begin{pmatrix}
  	\mathbf{0} & \mathbf{F} \\ \mathbf{G}  & \mathbf{0}
  \end{pmatrix} \begin{pmatrix}
  	\mathbf{X} \\ \mathbf{Y}
  \end{pmatrix}=ik\begin{pmatrix}
  	\mathbf{X} \\ \mathbf{Y}
  \end{pmatrix}.
\end{equation} 
by simply modifying the operators 
\begin{equation}
  \begin{aligned}
  	\mathbf{F}=\begin{pmatrix}
  	-\frac{1}{\gamma \tau} & -\frac{\gamma-2}{\gamma}\partial_{y} \\ 
  	\frac{\gamma-2}{\gamma}\partial_{y}+\frac{2\beta}{\gamma} & f_4
   \end{pmatrix},
    \ \
   \mathbf{G}=\begin{pmatrix}
   	\Omega^{2} c_r & \partial_{y} \\ -\partial_{y} & \frac{1}{\tau}
   \end{pmatrix}
  \end{aligned}
  \label{eq:F G}
\end{equation}
with 
\begin{equation*}
	f_4=-\Omega^{2} c_r -\frac{4\tau(\gamma-1)}{\gamma}\partial_{y^{2}}-\frac{4\alpha(\gamma-1)+4\beta \tau}{\gamma}\partial_{y},
\end{equation*}
where 
\begin{equation}
	\gamma=2(1-\nu)/(1-2\nu).
\end{equation}
It should be noticed that the dimensionless frequency is defined as $\Omega=\omega h/c_0$, with $c_0=1000\pi \ \text{m/s}$, such that the velocity in (\ref{eq:F G}) is $c_r=c_0^2/c_T^2$. 

There are three more parameters than the formulation of homogeneous plates
\begin{subequations}
\begin{equation}
	\tau(y)=\mu(y)/\mu_b,
\end{equation}
\begin{equation}
	\alpha=\partial_{y}\tau,
\end{equation}
\begin{equation}
	\beta(y)=\partial_{y}\gamma/\gamma,
\end{equation}
\end{subequations}
due to the variation of parameters along thickness direction.

\end{appendix}

\section*{References}

\bibliography{Bib_FGP_v2_new.bib}

\end{document}